\documentclass{emulateapj}

\shorttitle{The effect of $^{22}$Ne diffusion}
\shortauthors{Camisassa et al.}

\usepackage{natbib}
\begin{document}

\title{The  effect  of  $^{22}$Ne   diffusion  in  the  evolution  and
  pulsational  properties  of  white  dwarfs  with  solar  metallicity
  progenitors}

\author{Mar\'ia E. Camisassa,
        Leandro G. Althaus,
        Alejandro H. C\'orsico}
\affil{Facultad de Ciencias Astron\'omicas y Geof\'isicas,
       Universidad Nacional de La Plata,
       Paseo del Bosque s/n,
       1900 La Plata,
       Argentina}
\affil{Instituto de Astrof\'\i sica de La Plata,
       UNLP-CONICET,
       Paseo del Bosque s/n,
       1900 La Plata,
       Argentina}
\author{N\'uria Vinyoles, Aldo M. Serenelli, Jordi Isern}
\affil{Instituto de Ciencias del Espacio (CSIC),
       Carrer de Can Magrans s/n,
       08193, Cerdanyola del Vall\'es, 
       Spain}
\affil{Institute for Space Studies of Catalonia,
       c/Gran Capit\`a 2-4,
       Edif. Nexus 201,
       08034 Barcelona,
       Spain}
\author{Marcelo M. Miller Bertolami}
\affil{Max-Planck-Institut f\"ur Astrophysik,
       Karl-Schwarzschild-Str. 1,
       85748, Garching,
       Germany}
\affil{Instituto de Astrof\'\i sica de La Plata,
       UNLP-CONICET,
       Paseo del Bosque s/n,
       1900 La Plata,
       Argentina}
\and
\author{Enrique Garc\'\i a--Berro}
\affil{Departament de F\'\i sica,
       Universitat Polit\`ecnica de Catalunya,
       c/Esteve Terrades 5,
       08860 Castelldefels, 
       Spain}
\affil{Institute for Space Studies of Catalonia,
       c/Gran Capit\`a 2-4,
       Edif. Nexus 201,
       08034 Barcelona,
       Spain}
\email{camisassa@fcaglp.unlp.edu.ar}

\begin{abstract}
Because of the large neutron excess of $^{22}$Ne, this isotope rapidly
sediments in the interior of  the white dwarfs.  This process releases
an additional amount of energy, thus delaying the cooling times of the
white dwarf. This influences the ages of different stellar populations
derived   using  white   dwarf   cosmochronology.   Furthermore,   the
overabundance of $^{22}$Ne in the  inner regions of the star, modifies
the  Brunt-V\"ais\"al\"a  frequency,  thus  altering  the  pulsational
properties of  these stars.  In  this work,  we discuss the  impact of
$^{22}$Ne  sedimentation   in  white   dwarfs  resulting   from  Solar
metallicity   progenitors  ($Z=0.02$).    We  performed   evolutionary
calculations of white  dwarfs of masses $0.528$,  $0.576$, $0.657$ and
$0.833$  M$_{\sun}$, derived  from full  evolutionary computations  of
their progenitor stars, starting at the Zero Age Main Sequence all the
way through central hydrogen and helium burning, thermally-pulsing AGB
and post-AGB phases.   Our computations show that  at low luminosities
($\log(L/L_{\sun})\la  -4.25$),  $^{22}$Ne  sedimentation  delays  the
cooling of  white dwarfs with  Solar metallicity progenitors  by about
1~Gyr.   Additionally,  we  studied   the  consequences  of  $^{22}$Ne
sedimentation on  the pulsational properties of  ZZ~Ceti white dwarfs.
We  find  that  $^{22}$Ne  sedimentation induces  differences  in  the
periods  of   these  stars  larger  than   the  present  observational
uncertainties, particularly in more massive white dwarfs.
\end{abstract}


\section{Introduction}
\label{introduction}

White dwarf stars are the  most common end-point of stellar evolution.
They  provide a  wealth of  information about  the evolution  of their
progenitor stars, and the physical processes occurring in stars.  They
also provide valuable information about  the star formation history of
the  Solar neighborhood,  and  allow  us to  study  the properties  of
various   stellar   populations  --   see   \cite{2008PASP..120.1043F,
2008ARA&A..46..157W}, and  \cite{2010A&ARv..18..471A} for  reviews. In
particular, white  dwarfs are  used as accurate  age indicators  for a
wide variety of Galactic populations, including the disk, and open and
globular      clusters      --     see      \cite{2009ApJ...693L...6W,
2010Natur.465..194G,     2011ApJ...730...35J,     2013A&A...549A.102B,
2013Natur.500...51H}  and \cite{2015A&A...581A..90T}  for some  recent
applications.

The use of white dwarfs as reliable and precise clocks to date stellar
populations  has prompted  the  computation of  detailed and  complete
evolutionary  models for  these  stars, taking  into  account all  the
relevant sources and sinks of  energy, and the evolutionary history of
progenitor   stars  \citep{2010ApJ...717..183R,   2010ApJ...716.1241S,
2012A&A...537A..33A,  2015A&A...576A...9A}.  The  computation of  such
models requires  a detailed knowledge  of the main  physical processes
responsible for  their evolution. Among these  processes, and relevant
for the present paper, is the slow gravitational settling of $^{22}$Ne
in the  liquid phase,  that has  been shown  to strongly  decrease the
cooling rate  of white  dwarfs resulting from  high-metallicity ($Z\ga
0.03$)   progenitors  \citep{2008ApJ...677..473G,2010ApJ...719..612A}.
$^{22}$Ne is the  most abundant impurity present  in the carbon-oxygen
interiors of typical white dwarfs, and is the result of helium burning
of the $^{14}$N built up during  the CNO cycle. In particular, the two
additional  neutrons present  in  the $^{22}$Ne  nucleus (relative  to
$A=2Z$)  results in  a net  downward  gravitational force  and a  slow
settling of $^{22}$Ne in the liquid  regions towards the center of the
white  dwarf  \citep{1992A&A...257..534B}.    The  role  of  $^{22}$Ne
sedimentation  in the  energetics  of crystallizing  white dwarfs  was
first  addressed  by  \cite{1991A&A...241L..29I},  and  quantitatively
explored     later     by      \cite     {2002ApJ...580.1077D}     and
\cite{2010ApJ...719..612A}.   These  studies   showed  that  $^{22}$Ne
sedimentation releases  substantial energy  to modify  appreciably the
cooling  of massive  white  dwarfs, delaying  the  evolution by  about
$10^9$ yrs  at low  luminosities ($\log  L/L_\odot\lesssim-4.5$).  The
occurrence of  this process in the  interior of cool white  dwarfs has
been  shown  to be  a  key  factor  in  solving the  longstanding  age
discrepancy      of      the     metal-rich      cluster      NGC~6791
\citep{2010Natur.465..194G}

The       studies        of       \cite{2008ApJ...677..473G}       and
\cite{2010ApJ...719..612A}  revealed the  necessity  of including  the
gravitational  energy  released  by  $^{22}$Ne  sedimentation  in  the
calculations of  detailed white dwarf cooling  sequences. Motivated by
these findings, in  this paper we investigate the  impact of $^{22}$Ne
sedimentation on the cooling age  of white dwarfs resulting from Solar
metallicity progenitors.  Our aim is to provide an accurate assessment
of the differences in the cooling ages when this process is neglected.
The results  presented here  improve our previous  works in  two ways.
First,  the  calculations  presented here  include  realistic  initial
$^{22}$Ne  abundance  distributions  in   the  white  dwarf  interior,
resulting  from  the evolutionary  history  of  progenitor stars.   In
particular, we compute the full evolution of progenitor stars starting
from the  ZAMS all the way  through the phases of  hydrogen and helium
core  burning  and  the  thermally  pulsing  asymptotic  giant  branch
(AGB). In this way we obtained realistic initial $^{22}$Ne profiles at
the  beginning  of the  white  dwarf  cooling sequence,  necessary  to
accurately  compute   the  energy  released  by   the  slow  $^{22}$Ne
sedimentation  along  the  entire  white  dwarf  cooling  track.  And,
secondly, the physical description  of $^{22}$Ne diffusion in strongly
coupled plasma  mixtures is substantially improved.   Specifically, in
our  calculations  we  employ   the  new  diffusion  coefficients  for
$^{22}$Ne  based on  the molecular  dynamics simulations  of $^{12}$C,
$^{16}$O, and $^{22}$Ne mixtures of \cite{2010PhRvE..82f6401H}.  These
diffusion coefficients  are now accurately known.  Thus, the remaining
uncertainties  in their  specific values  should not  be relevant  for
white dwarf evolutionary calculations.

We extend  the scope  of the  paper by  exploring the  consequences of
$^{22}$Ne sedimentation  for the  adiabatic pulsational  properties of
ZZ~Ceti  stars.   To this  end,  we  perform an  adiabatic,  nonradial
pulsation analysis  of $g$-modes. The  results presented here  are the
first ones  in showing  the impact of  $^{22}$Ne sedimentation  on the
expected  spectrum  of  pulsation  periods of  evolving  ZZ~Ceti  star
models.

\begin{table}
\caption{Basic model properties for our sequences. We list the mass at
  the  zero-age  main  sequence  ($M_{\rm ZAMS}$),  the  mass  of  the
  resulting white  dwarf ($M_{\rm WD}$),  the age at the  beginning of
  white dwarf  cooling sequence, defined  at the moment when  the star
  reaches  the maximum  effective temperature,  $t_{\rm pre-WD}$,  and
  surface carbon to  oxygen ratio at the beginning of  the white dwarf
  stage (C/O).}  \centering
\begin{tabular}{lccc}
\tableline
\tableline
$M_{\rm ZAMS}\, (M_{\sun})$ &$M_{\rm WD}\, (M_{\sun})$ &$t_{\rm pre-WD}$~(Gyr) & C/O \\
\tableline
1.0 & 0.528 & 11.813 & 0.3009 \\
1.5 & 0.576 &  2.820 & 1.6488\\
3.0 & 0.657 &  0.443 & 1.1434 \\
4.0 & 0.833 &  0.196 & 4.0894 \\
\tableline
\tableline
\end{tabular}
\label{tabla1}
\end{table}

\section{Numerical setup and input physics}
\label{code}

The evolutionary calculations  discussed in this paper  were done with
an updated version of the  {\tt LPCODE} stellar evolutionary code, see
\cite{2005A&A...441..689A}   and  references   therein.   This   is  a
well-tested and calibrated code that has  been amply used in the study
of different aspects of the evolution of low-mass stars. Particularly,
it has  been used to compute  very accurate and realistic  white dwarf
models         \citep{2008A&A...491..253M,        2010Natur.465..194G,
2010ApJ...717..897A,     2010ApJ...717..183R,     2011ApJ...743L..33M,
2011A&A...533A.139W,     2012MNRAS.424.2792C,     2013A&A...557A..19A,
2015A&A...576A...9A}.   A recent  test comparing  {\tt LPCODE}  with a
different white  dwarf evolutionary  code shows that  uncertainties in
the white dwarf cooling ages  that result from the different numerical
implementations of the  stellar evolution equations are  less than 2\%
\citep{2013A&A...555A..96S}.   A  detailed  description of  the  input
physics and numerical procedures can be found in these works.  In this
section we briefly summarize the main input physics for this work.

\begin{figure}
\centering
\includegraphics[clip,width=0.9\columnwidth]{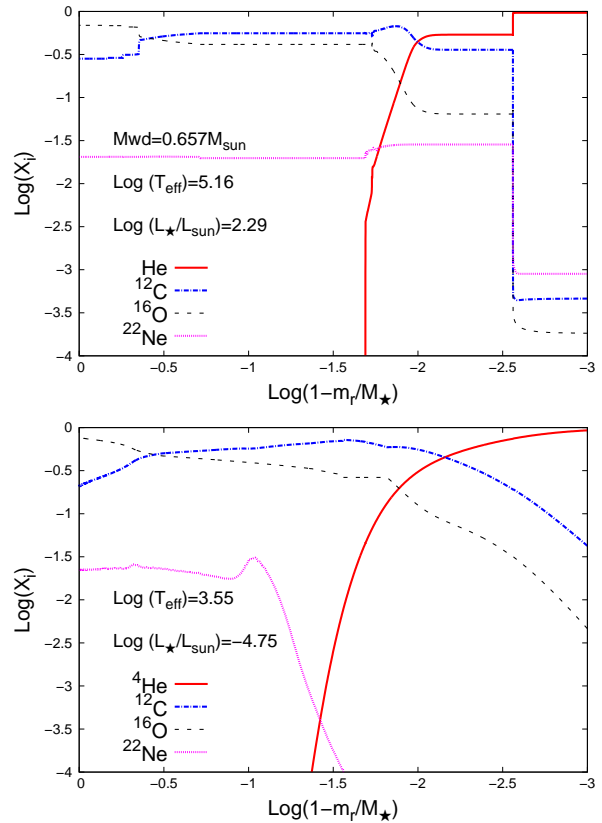}
\caption{Logarithmic mass  fractions of  the elements in  the $0.657\,
  M_{\sun}$ white dwarf  model. The top panel shows  the abundances at
  the beginning of the white dwarf cooling sequences, while the bottom
  panel shows the final abundances.}
\label{Fig:profiles}
\end{figure}

In this  work initial white  dwarf models  have been derived  from the
full  evolutionary history  of their  progenitor stars,  including the
central hydrogen and helium burning, the thermally-pulsing AGB and the
post AGB.  The initial metallicity  of the  sequences has been  set to
$Z=0.02$.  The initial masses of our sequences and the resulting white
dwarf  masses are  listed  in Table  \ref{tabla1},  together with  the
pre-white dwarf age and the carbon-to-oxygen ratio at the beginning of
the white dwarf stage, before  gravitational settling modifies it. The
evolution of the  models before the white dwarf stage  is described in
detail in \cite{2015arXiv151204129M}, to which the reader is referred.

In  contrast with  our previous  studies  of the  impact of  $^{22}$Ne
sedimentation  on  white  dwarf  evolution,   in  this  work  we  have
considered realistic  initial $^{22}$Ne  profiles at the  beginning of
the  cooling track.   Relevant to  this aspect,  we mention  that {\tt
LPCODE}  considers  a   simultaneous  treatment  of  non-instantaneous
mixing(and extra-mixing  if present), element diffusion  (chemical and
thermal diffusion plus gravitational  settling) and nuclear burning of
elements.  The code  treats convective boundary mixing  as a diffusion
process by assuming that  mixing velocities decay exponentially beyond
each                        convective                        boundary
\citep{1996A&A...313..497F,1997A&A...324L..81H}.    Specifically,   we
assumed  a diffusion  coefficient  $D_{\rm OV}=D_{\rm  O} \exp  (-2z/f
H_{\rm p})$,  where $H_{\rm p}$  is the  pressure scale height  at the
convective  boundary,  $D_{\rm O}$  is  the  diffusion coefficient  of
unstable  regions close  to the  convective boundary,  and $z$  is the
geometric  distance   from  the  edge  of   the  convective  boundary.
Convective  boundary  mixing  has  been  considered  during  the  core
hydrogen and helium  burning phases and in  the thermally-pulsing AGB.
The convective boundary  mixing efficiency $f$ has  been calibrated to
different  values  in  different  convective boundaries  in  order  to
reproduce different  observables during  the pre-white dwarf  phase --
see \cite{2015arXiv151204129M}  for details. Radiative  and conductive
opacities are taken from the OPAL database \citep{1996ApJ...464..943I}
and   from   \cite{2007ApJ...661.1094C},    respectively.    For   the
low-temperature regime  we used  the molecular opacities  with varying
carbon-to-oxygen  ratios of  \cite{2005ApJ...623..585F}, presented  in
\cite{2009A&A...508.1343W}.   These  opacities  are  necessary  for  a
realistic  treatment of  the evolution  of the  progenitor during  the
thermally-pulsing  AGB phase.   At low  effective temperatures  in the
white dwarf regime, outer boundary  conditions for the evolving models
are      derived       from      non-grey       model      atmospheres
\citep{2012A&A...546A.119R}.   Latent  heat  and phase  separation  of
carbon and oxygen due to  crystallization have been included as energy
sources, using the phase diagram of \cite{2010PhRvL.104w1101H}.

\begin{figure}
\centering
\includegraphics[clip,width=0.9\columnwidth]{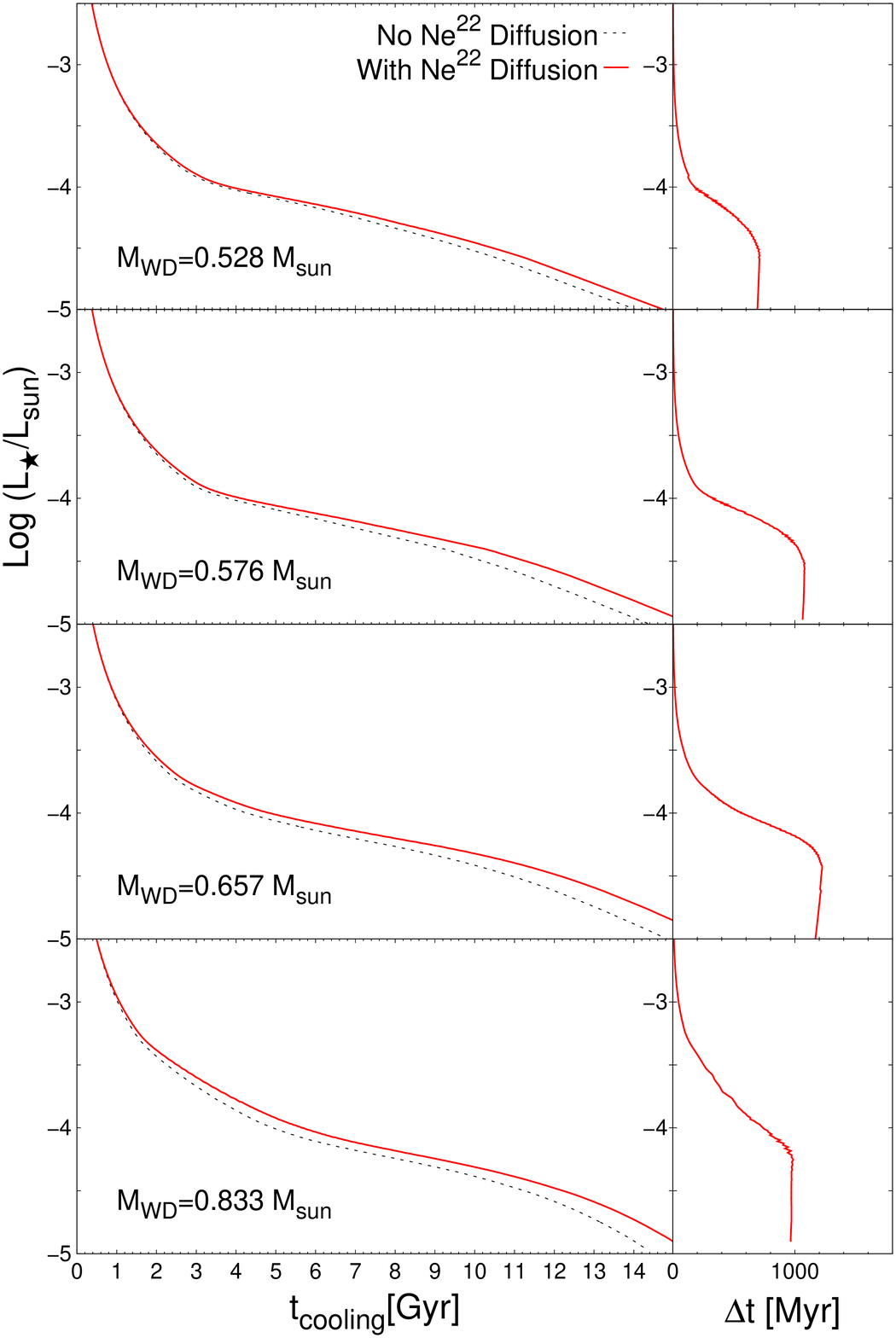}
\caption{Impact  of  $^{22}$Ne  sedimentation on  the  cooling  times,
  defined as  the time  since the star  reaches its  maximum effective
  temperature,  for our  four  sequences. The  dashed  line shows  the
  evolution  when $^{22}$Ne  sedimentation is  disregarded, while  the
  solid  red  line  shows  the  evolution for  the  white  dwarf  when
  $^{22}$Ne sedimentation  is included.  Time delays  in Myr  are also
  shown in the right panels.}
\label{Fig:delays}
\end{figure}

Another improvement with  respect to our previous studies,  is that we
have  included in  this  work  the most  recent  determination of  the
$^{22}$Ne diffusion coefficients  \citep{2010PhRvE..82f6401H}. We have
also improved  our numerical treatment for  $^{22}$Ne sedimentation by
assuming that $^{22}$Ne diffuses in a one component plasma background,
consisting of a fictitious element with atomic weight ($A$) and atomic
number ($Z$).  In contrast with  our previous treatment, here  $A$ and
$Z$ of the  fictitious element are defined by the  average $A$ and $Z$
in  each layer,  and not  assumed constant  through different  layers.
Finally, the  energy released  by $^{22}$Ne  sedimentation as  well as
from  crystallization   is  included   locally  in  the   equation  of
luminosity,       following       \cite{2008ApJ...677..473G}       and
\cite{2010ApJ...719..612A}. The calculations presented here constitute
the first grid of Solar-metallicity white dwarf evolutionary sequences
that include  the effects of  $^{22}$Ne sedimentation in  a consistent
way.   Finally, for  the sake  of comparison,  we have  also performed
additional evolutionary calculations  in which $^{22}$Ne sedimentation
has been suppressed.

\begin{figure}
\centering
\includegraphics[clip,width=0.9\columnwidth]{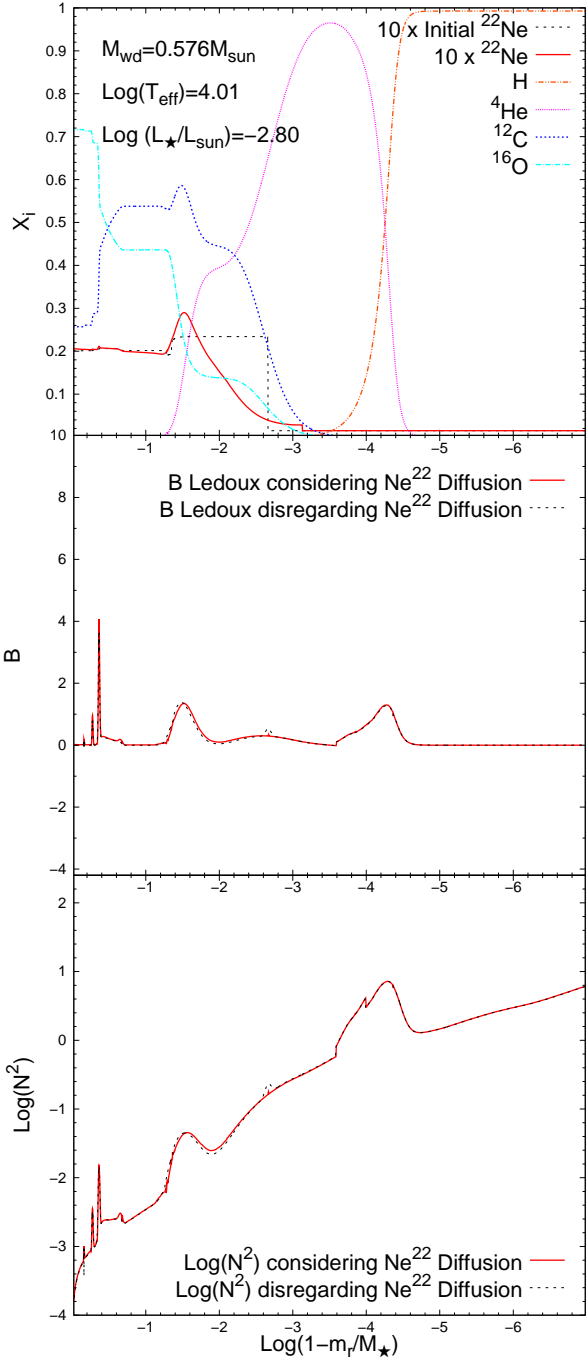}
\caption{Top  panel: chemical  abundance  distribution  of a  selected
  $0.576\,  M_{\sun}$   white  dwarf  model  at   $\log(T_{\rm  eff})=
  4.01$. For  comparison purposes, the initial  $^{22}$Ne distribution
  at the beginning  of the cooling track is also  shown. Middle panel:
  Ledoux term for  the same model. Solid (dotted)  line corresponds to
  the    case    when     $^{22}$Ne    sedimentation    is    included
  (disregarded).    Bottom   panel:    run   of    the   corresponding
  Brunt-V\"ais\"al\"a frequency.}
\label{Fig:fvb_057}
\end{figure}

In order to improve upon previous  works we also analyze the impact of
$^{22}$Ne  sedimentation  in  the predicted  pulsational  spectrum  of
ZZ~Ceti stars.  For this we employ  the adiabatic version of  the {\tt
LP-PUL}     pulsation     code      described     in     detail     in
\citet{2006A&A...454..863C},  that  is  coupled to  the  {\tt  LPCODE}
evolutionary  code. The  {\tt LP-PUL}  pulsation  code is  based on  a
general Newton-Raphson technique that solves the full fourth-order set
of real equations and boundary conditions governing linear, adiabatic,
nonradial  pulsations  following   the  dimensionless  formulation  of
\citet{1971AcA....21..289D}  --  see \cite{1989nos..book.....U}.   The
prescription we  follow to assess  the run of  the Brunt-V\"ais\"al\"a
frequency  ($N$) for  a  degenerate environment  typical  of the  deep
interior  of  a  white  dwarf is  the  so-called  ``Ledoux  Modified''
treatment \citep{1990ApJS...72..335T}:

\begin{figure}
\centering
\includegraphics[clip,width=0.9\columnwidth]{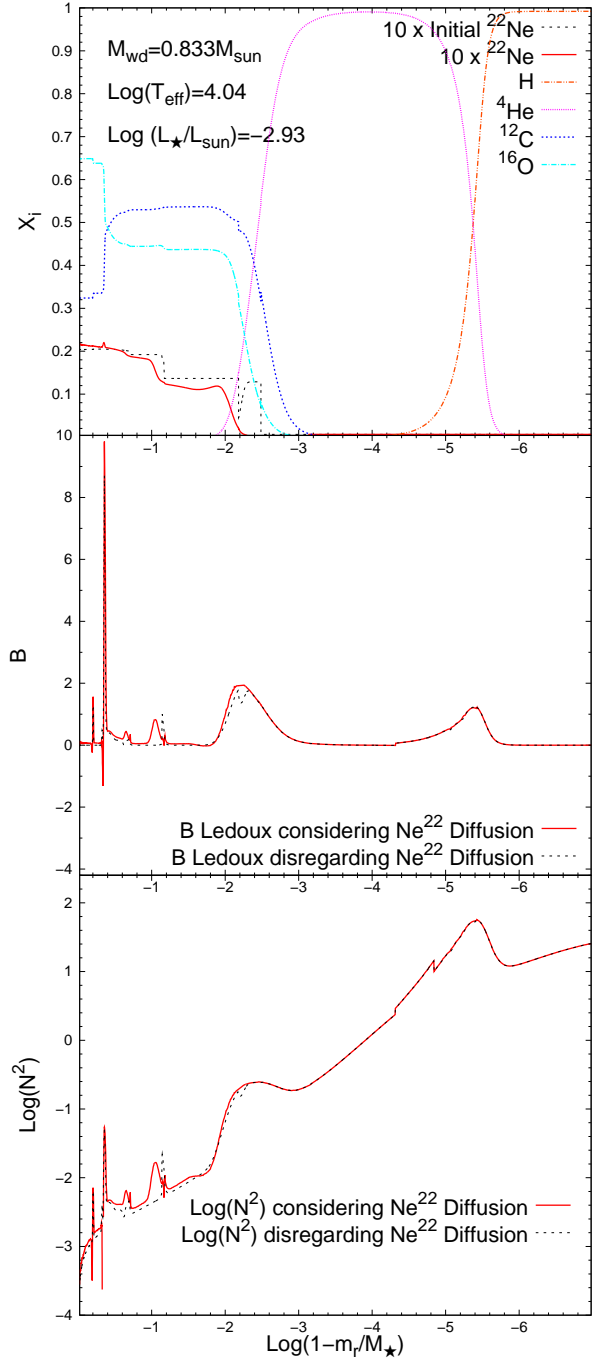}
\caption  {Same as  Fig.~\ref{Fig:fvb_057}  for  a $0.833\,  M_{\sun}$
  white dwarf model at $\log(T_{\rm eff})= 4.04$.}
\label{Fig:fvb_083}
\end{figure}

\begin{equation}
N^2= \frac{g^2 \rho}{P}\frac{\chi_{\rm T}}{\chi_{\rho}}
\left[\nabla_{\rm ad}- \nabla + B\right],
\label{bv}
\end{equation}

\noindent where the compressibilities are defined as

\begin{equation}
\chi_{\rho}= \left(\frac{\partial \ln P}{\partial \ln \rho}\right)_{{\rm T}, \{\rm X_i\}}\ \ \
\chi_{\rm T}= \left(\frac{\partial \ln P}{\partial \ln T}\right)_{\rho, \{\rm X_i\}}.
\end{equation}

\noindent The Ledoux term $B$ is computed as \citep{1990ApJS...72..335T}:

\begin{equation}
B= -\frac{1}{\chi_{\rm T}} \sum_1^{M-1} \chi_{\rm X_i} \frac{d\ln X_i}{d\ln P}, 
\label{BLedoux}
\end{equation}

\noindent where:

\begin{equation}
\chi_{\rm X_i}= \left(\frac{\partial \ln P}{\partial \ln X_i}\right)_{\rho, {\rm T}, 
\{\rm X_{j \neq i}\}}.
\end{equation}

\section{Results}

\label{results}

\subsection{Impact on white dwarf evolution} 
\label{evres}

The inner chemical  abundance distribution in terms of  the outer mass
fraction  at  the beginning  of  the  cooling  track of  our  $0.657\,
M_{\sun}$  white  dwarf  sequence  is  shown in  the  upper  panel  of
Fig.~\ref{Fig:profiles}.     This    figure   shows    the    chemical
stratification as given by the  evolutionary history of the progenitor
star.  In  particular, the $^{22}$Ne  abundance by mass is  about 0.02
throughout the  innermost region of  the white dwarf.   This abundance
was mostly built  up during the helium core burning  phase and results
from   helium   burning   on   $^{14}$N  via   the   reactions   \mbox
{$^{14}$N($\alpha,        \gamma$)$^{18}$F($\beta^+$)$^{18}$O($\alpha,
\gamma$)$^{22}$Ne}.   It is  also worth  noting the  abrupt change  of
$^{22}$Ne abundance  at $\log  (1-m_r/M_\star) \sim -2.5$,  just below
the pure helium buffer.  This abrupt change in all chemical abundances
shown  in the  figure marks  the  extent reached  by the  pulse-driven
convection zone during  the last thermal pulse on  the AGB experienced
by the progenitor star.

\begin{table*}
\caption  {White  dwarf  cooling  ages for  sequences  with  $^{22}$Ne
  diffusion  and  their  corresponding  time delays  compared  to  the
  sequences in which this process is disregarded.}  \centering
\begin{tabular}{lcccc}
\tableline
\tableline
\multicolumn{1}{c}{$-\log(L/L_\sun)$} & \multicolumn{4}{c}{$t_{\rm cool}$~(Gyr)}\\
\tableline
\multicolumn{1}{c}{} & $0.528\, M_{\sun}$ & $0.576\, M_{\sun}$ & $0.657\, M_{\sun}$ & $0.833\, M_{\sun}$\\
\tableline
1.0 &  0.02  &  0.02  &  0.01 &  0.01 \\
2.0 &  0.16  &  0.16  &  0.17 &  0.21 \\	
3.0 &  0.77  &  0.80  &  0.86 &  1.06 \\		
4.0 &  3.92  &  4.15  &  4.83 &  5.65 \\
4.5 & 10.45  & 11.33  & 12.10 & 12.13 \\
5.0 & 14.68  & 15.29  & 16.01 & 15.52 \\
\tableline
\multicolumn{1}{c}{} & \multicolumn{4}{c}{$\delta t$~(Gyr)}\\
\tableline
3.0 & 0.006 & 0.009 & 0.015 & 0.046 \\ 
4.0 & 0.18  &  0.34 &  0.56 &  0.74 \\ 
4.5 & 0.69  &  1.07 &  1.22 &  0.97 \\
\tableline
\tableline
\end{tabular}
\label{tabladeltat}
\end{table*}

The bottom  panel of  Fig.~\ref{Fig:profiles} shows  the corresponding
chemical  stratification at  the end  of the  cooling track,  when the
surface luminosity  is $\log(L/\rm L_\sun)=-4.75$.  Note  that element
diffusion  has   markedly  changed  the  initial   chemical  abundance
distribution. In  particular, chemical transitions have  been smoothed
out by the  diffusion processes. Note also that  element diffusion has
depleted $^{22}$Ne in  the outer regions of the core,  and enhanced it
in the  central region of  the white  dwarf.  This takes  place during
those stages  of the evolution  during which  the white dwarf  core is
liquid, and is more relevant for  massive white dwarfs, owing to their
larger gravities.

\cite{2002ApJ...580.1077D,2008ApJ...677..473G}                     and
\cite{2010ApJ...719..612A}  have  convincingly  shown  that  the  slow
$^{22}$Ne  sedimentation process  releases enough  energy to  strongly
alter  the evolution  of white  dwarfs characterized  by a  high metal
content  in their  interiors.  Here,  we find  that cooling  times are
substantially  modified  in  the  case   of  white  dwarf  with  solar
metallicity    progenitors.     This     can    be    inferred    from
Fig.~\ref{Fig:delays}, that  shows the white dwarf  surface luminosity
as a function of age for  all our sequences.  The solid (dotted) lines
show  the  prediction  when   $^{22}$Ne  sedimentation  is  considered
(suppressed).  Time delays  are shown in the right  side panels.  Note
the  substantial   lengthening  of  the  evolutionary   times  at  low
luminosities.  At the luminosities typical of the cut-off of the white
dwarf luminosity  function, $\log  (L/\rm L_\sun) \approx  -4.5$, time
delays range from  0.7 to 1.2~Gyr, depending on the  stellar mass (see
Table \ref{tabladeltat}).  It should be mentioned that the oldest 
white dwarfs in the Galactic disk are likely low metallicity, so
the impact of $^{22}$ Ne sedimentation on those white dwarfs should be less 
relevant. Also note that massive white dwarfs have larger gravities,
and  therefore  $^{22}$Ne sedimentation  is  more  effective.  At  low
luminosities,  when  the  white dwarf  crystallizes,  $^{22}$Ne  stops
diffusing inwards  through the solid  phase, therefore the  time delay
stops  growing. This  is  reflected in  the curve  of  the right  side
panels, as it becomes constant after crystallization.  Time delays are
slightly shorter in the $0.833\, M_\sun$ white dwarf model than in the
$0.657\, M_\sun$  model, because more massive  white dwarf crystallize
at a  higher luminosity, preventing  $^{22}$Ne to keep  diffusing.  In
summary,  $^{22}$Ne sedimentation,  a process  not considered  in most
existing  grids of  white dwarf  evolutionary sequences  used to  date
stellar populations,  induces substantial delays in  the cooling times
of  white dwarfs  with  Solar metallicity  progenitors. Moreover,  the
lengthening of the cooling times is much larger than the uncertainties
arising from  current uncertainties in the  microphysics and numerical
inputs  \citep{2013A&A...555A..96S}.    A  more   detailed  comparison
between  model   cooling  sequences  and  the   observed  white  dwarf
luminosity function is thus desirable.   We defer these comparisons to
future work.

\subsection{Asteroseismological consequences} 
\label{ares}

Early estimates of the impact  of $^{22}$Ne diffusion on the pulsation
properties of  ZZ~Ceti stars  was done  by \cite{2002ApJ...580.1077D},
who found that this process could change the period of the high radial
order  $g-$modes by  about  1  $\%$. In  this  section,  we perform  a
detailed analysis  to provide a  reliable assessment of  the adiabatic
pulsation properties of our white  dwarf evolutionary models that take
into account $^{22}$Ne diffusion. We  compute pulsation periods in the
range  $100  \lesssim \Pi  \lesssim  3500$~s  corresponding to  dipole
($\ell=  1$)  $g$  modes.  Quantities  relevant  for  the  pulsational
properties of  our models  are shown  in Fig.~\ref{Fig:fvb_057}  for a
selected $0.576 \, M_{\sun}$ white  dwarf model at $\log(T_{\rm eff})=
4.01$.   For comparison  purposes,  we  show in  the  upper panel  the
chemical stratification of  the model. Note that at this  stage of the
evolution, which corresponds  to the domain of  ZZ~Ceti stars, element
diffusion has  markedly altered the initial  $^{22}$Ne distribution in
the  star. In  particular, it  has  strongly smoothed  out the  abrupt
change of the $^{22}$Ne abundance at $\log (1-m_r/M_\star) \sim -2.6$,
and it has  produced a bump in its abundance  at $\log (1-m_r/M_\star)
\sim  -1.5$.   This  bump  is  located  at  the  tail  of  the  helium
distribution, where  the average values  of $Z$ and $A$  grow inwards,
leading to a change in the  diffusion coefficients at those layers. In
particular,  the  diffusion  coefficients  decrease  inwards  at  that
interface, thus $^{22}$Ne ions tend  to accumulate. As the white dwarf
evolves, this bump diffuses inwards.

In the middle and bottom  panels of Fig.~\ref{Fig:fvb_057} we plot the
run  of  the  Ledoux  term  $B$  and  the  logarithm  of  the  squared
Brunt-V\"ais\"al\"a frequency, respectively, for  the same white dwarf
model.   We show  the  model including  Ne$^{22}$ sedimentation  using
solid lines, while  dotted lines are employed for the  model for which
Ne$^{22}$  sedimentation was  neglected.   As can  be seen,  $^{22}$Ne
sedimentation  barely affects  the  shape  of the  Brunt-V\"ais\"al\"a
frequency,  being almost unnoticeable in the plot.

\begin{figure}
\centering
\includegraphics[clip,width=0.9\columnwidth]{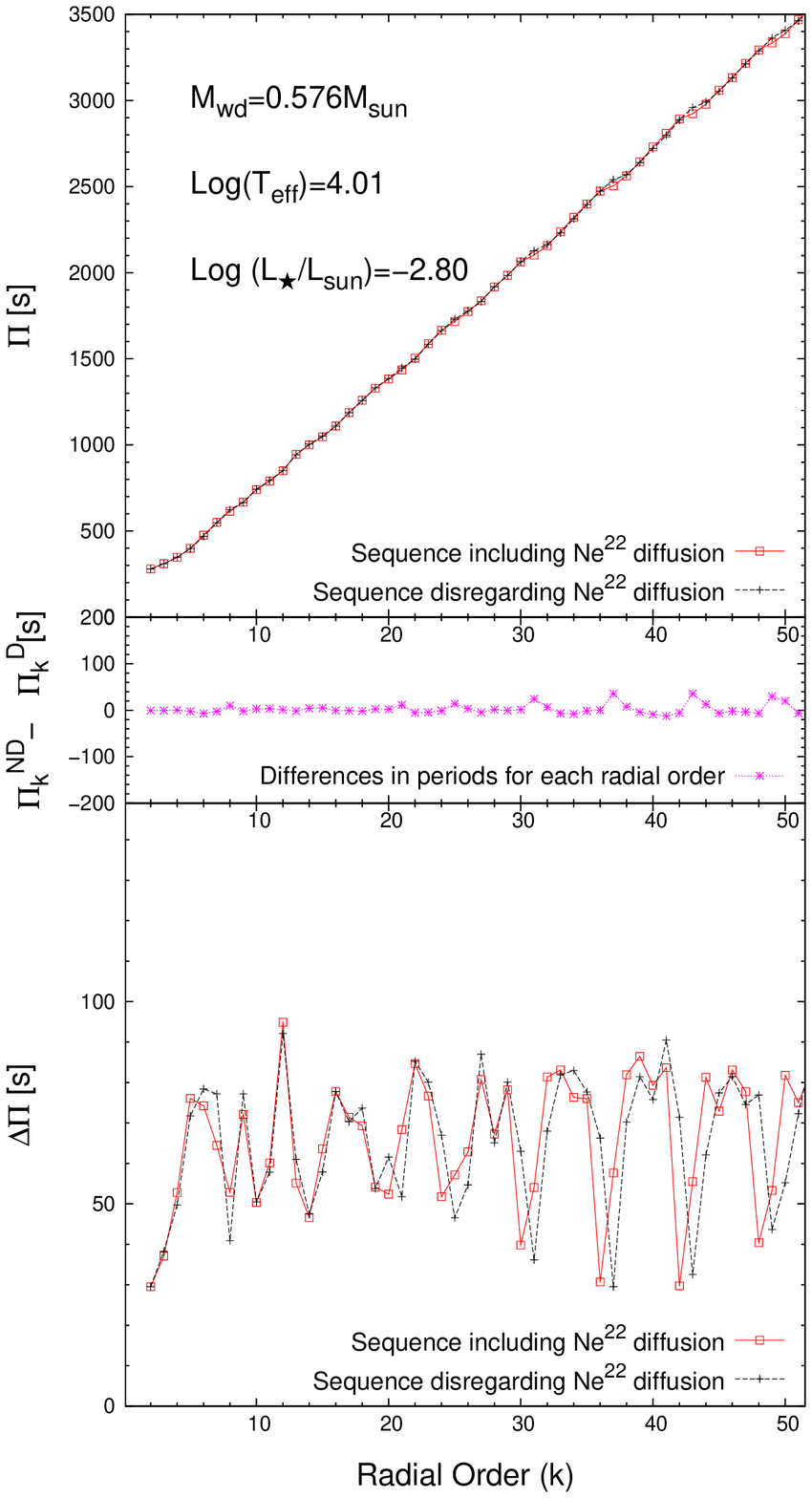}
\caption{Top,  middle   and  bottom  panel  show,   respectively,  the
  pulsation periods in  terms of the radial order  $k$, the difference
  in the  periods induced by  $^{22}$Ne sedimentation for  each radial
  order, and  the forward period  spacing. The square  (cross) symbols
  correspond  to the  case  in which  $^{22}$Ne  diffusion process  is
  included  (disregarded). Quantities  are  for  a $0.576\,  M_{\sun}$
  white dwarf model at $\log(T_{\rm eff})= 4.01$.}
\label{Fig:periods_057}
\end{figure}

\begin{figure}
\centering
\includegraphics[clip,width=0.9\columnwidth]{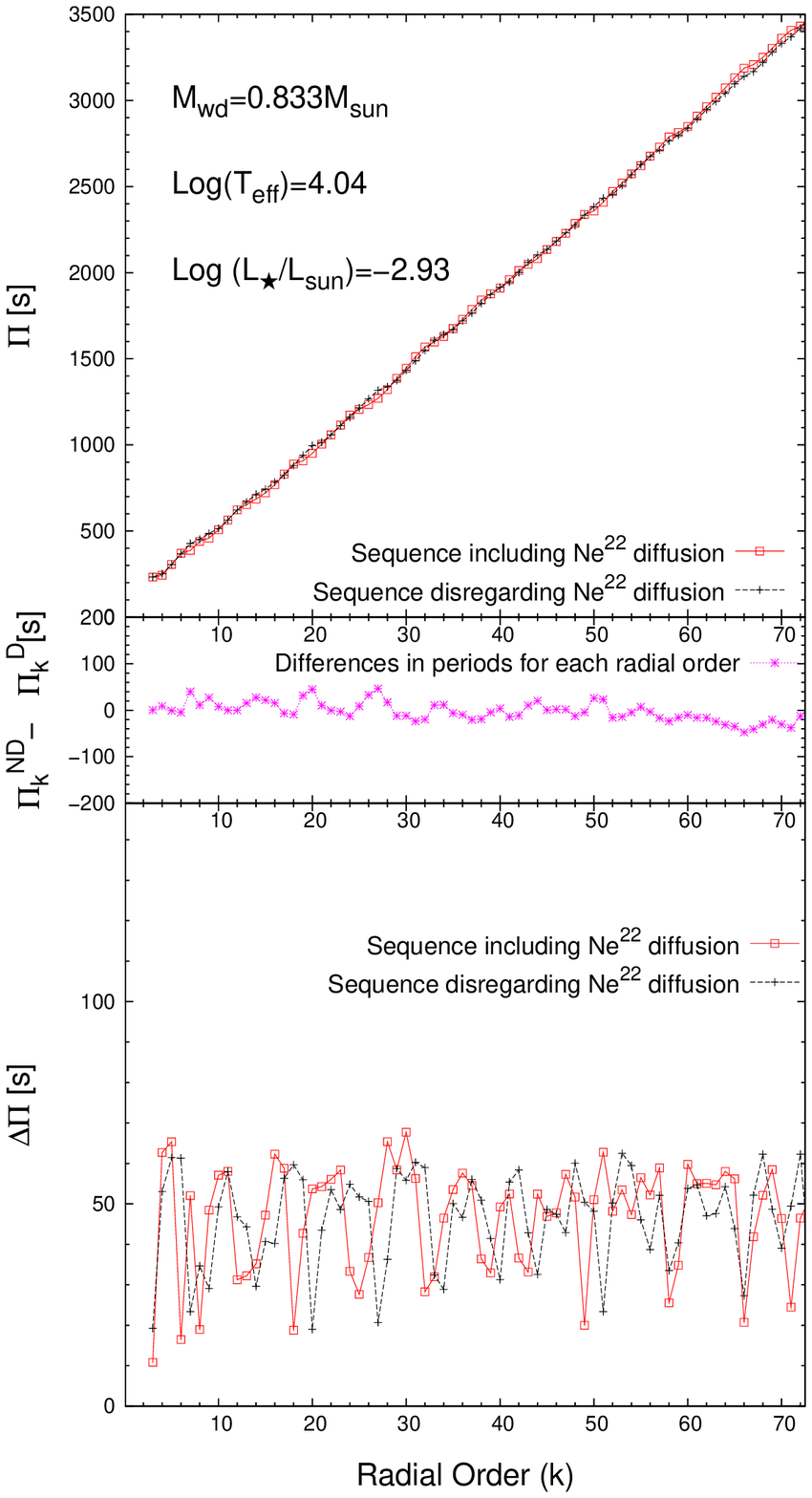}
\caption{Same  as  Fig.~\ref{Fig:periods_057},   but  for  a  $0.833\,
  M_{\sun}$ white dwarf at $\log(T_{\rm eff})= 4.04$.}
\label{Fig:periods_083}
\end{figure}

Because  of  the two  additional  neutrons  of the  $^{22}$Ne  nucleus
(relative  to $A=2\  Z$ nuclei),  changes in  the $^{22}$Ne  abundance
directly translate  into appreciable  changes in  the pressure  of the
degenerate electron gas.   By the time evolution has  proceeded to the
domain of the ZZ~Ceti stars, $^{22}$Ne has diffused toward the central
regions of the star, so the outer layers of the core have already been
depleted of this element. This behavior  turns out to be markedly more
noticeable for larger gravities.  This can be appreciated in the upper
panel of  Fig.~\ref{Fig:fvb_083}, which  shows the  chemical abundance
distribution of  a selected  $0.833\, M_{\sun}$  white dwarf  model at
$\log(T_{\rm eff})= 4.04$.   The consequences for the  Ledoux term and
the Brunt-V\"ais\"al\"a frequency can be  inferred from the middle and
bottom   panels,   respectively.    Note  in   particular   that   the
overabundance of $^{22}$Ne  in the inner regions of the  star leads to
an increase of  the density and therefore to a  global increase in the
Brunt-V\"ais\"al\"a frequency in those regions.

The  precise  shape  of   the  Brunt-V\"ais\"al\"a  frequency  largely
determines the structure  of the $g$-mode period  spectrum. Since this
quantity  is modified  by the  overabundance of  the $^{22}$Ne  in the
inner regions of  the star, particularly in massive  white dwarfs, the
value of the  pulsation periods are thus expected to  be affected when
sedimentation of $^{22}$Ne is allowed to operate. This is demonstrated
with the help of Figs.~\ref{Fig:periods_057} and \ref{Fig:periods_083}
for the  same models  analyzed in the  previous paragraph.   The upper
panel of each figure shows the $\ell=1$ pulsation periods, $\Pi$, as a
function  of  the radial  order  $k$.  The  middle and  bottom  panels
illustrate, respectively, the difference in the periods induced by the
sedimentation  process of  $^{22}$Ne, $\Pi_k^{\rm  ND}-\Pi_k^{\rm D}$,
and   the   forward   period    spacing,   $\Delta   \Pi_k$   ($\equiv
\Pi_{k+1}-\Pi_k$). The squared symbols (in red) correspond to the case
when $^{22}$Ne sedimentation  is considered and the  cross symbols (in
black) for the case in which  diffusion is disregarded. In the case of
the low-mass white dwarf model,  the impact of $^{22}$Ne sedimentation
on  the pulsation  periods  is  small and  barely  appreciable in  the
plot.  The  period   differences,  $|\Pi_k^{\rm  ND}-\Pi_k^{\rm  D}|$,
between the  models with  and without  $^{22}$Ne sedimentation  in the
range of  the periods observed in  ZZ~Ceti stars (from 100  to 1500~s)
are on  average $\sim 3$~s,  reaching values  as high as  $\sim 11$~s.
Although these differences are not large, they are still comparable to
 the typical average residual in model fits
to the observed periods of pulsating white dwarfs ($1-3$~s)
Finally,  the  forward  period spacing  (which  is  extremely
sensitive  to  the  precise  shape  of  the  chemical  profiles)  also
experiences a  noticeable change when  we take into  account $^{22}$Ne
sedimentation, as it is displayed in  the bottom panel of this figure.
In summary, the effects of  $^{22}$Ne sedimentation on the pulsational
properties  of our  model are  appreciable  for modes  with very  long
periods (high radial orders). However, we do not expect a large impact
on the $g$-mode  period spectrum of average-mass ZZ~Ceti  stars in the
range of periods typically observed in these pulsating white dwarfs.

The situation  is quite different for  the case of more  massive white
dwarfs, where $^{22}$Ne sedimentation becomes much more relevant. This
can be  deduced from  Fig.~\ref{Fig:periods_083} for a  selected white
dwarf model of our most massive  sequence.  Now the differences in the
theoretical periods that result from including $^{22}$Ne sedimentation
are on average $\sim 15$~s, with  values as high as $\sim 47$~s. These
differences are, by far, much  larger than the
 typical average residual in model fits to the observed periods of 
pulsating white dwarfs. The
impact of  considering $^{22}$Ne  sedimentation on the  forward period
spacing is also  noteworthy for this massive white  dwarf model.  Note
that the effects are larger for higher radial orders (long periods).
 Although our results show that $^{22}$Ne sedimentation affects the 
pulsation periods, it could still be difficult to infer its presence
 from a particular signature in the pulsation spectrum, since these 
differences may be mimicked by other changes in the models.

The rate  of period change,  $d\Pi/dt$, which depends on  the chemical
composition of the core, is  also affected by $^{22}$Ne sedimentation.
This quantity is  related to the rate of change  in the temperature of
the isothermal core ($\dot{T}$) through the equation:
\begin{equation}
\label{Pdot}
 \frac{\dot{P}}{P}=-a  \frac{\dot{T}}{T}+ b  \frac{\dot{R}_\star}{R_\star}
\end{equation}
where $a$ and $b$ are dimensionless constants of order unity depending
on  the equation  of state,  thicknesses  of the  hydrogen and  helium
layers,    chemical    composition,    amongst    others    parameters
\citep{1983Natur.303..781W}.   Usually,   for    white   dwarfs   with
hydrogen-rich atmospheres in the  ZZ~Ceti instability strip the second
term of the right hand side of equation \ref{Pdot} is negligible.  The
impact of considering  $^{22}$Ne sedimentation on the  rates of period
change, at $\log(T_{\rm  eff})\sim 4$, is larger for  the more massive
white dwarf model ($0.833\, M_{\sun}$) than for the $0.576\, M_{\sun}$
model.   In the  case  of  the more  massive  white  dwarf model  with
$\log(T_{\rm  eff})=  4.04$,  the  temporal  rates  of  period  change
obtained  by  the  calculations  are,   on  average,  $\sim  3  \times
10^{-15}$~s/s  larger   for  the  sequences  that   neglect  $^{22}$Ne
diffusion.   This result  was somehow  expected since  these sequences
cool faster  than the ones  that include $^{22}$Ne diffusion.   In the
case  of  the $0.576~M_{\sun}$  white  dwarf  model with  $\log(T_{\rm
eff})= 4.01$, no  appreciable difference, on average,  in the temporal
rates of  period change induced  by $^{22}$Ne diffusion  was obtained.
At this point  in the evolution ($\log(L/\rm  L_\sun)=-2.8$), the rate
of  change in  the temperature  of the  isothermal core  is still  not
affected    by    the    $^{22}$Ne    sedimentation    process    (see
\ref{Fig:delays}).  Thus,  no difference  in $d\Pi/dt$ is  expected to
occur as a consequence of this process.

\section{Summary and conclusions}
\label{conclusions}

The sedimentation of $^{22}$Ne is  a well established physical process
that has been shown to be  an important source of gravitational energy
during  the  cooling  of   white  dwarfs  descending  from  metal-rich
progenitor              stars              \citep{2002ApJ...580.1077D,
2008ApJ...677..473G,2010ApJ...719..612A}.  The  reason behind  this is
that  the  neutron excess  that  characterizes  the $^{22}$Ne  nucleus
yields a net  downward force and, consequently, in  the liquid regions
of the star $^{22}$Ne slowly diffuses  towards the center of the white
dwarf \citep{1992A&A...257..534B}.

Observational evidence  for the occurrence of  $^{22}$Ne sedimentation
in  the  liquid  interior  of   white  dwarfs  has  been  provided  by
\cite{2010Natur.465..194G}.  However,  most  of the  existing  cooling
sequences       do  not  take  this process  into  account.  The  only
exceptions   are   the   works   of   \cite{2008ApJ...677..473G}   and
\cite{2010ApJ...719..612A},  who  calculated   grids  of  evolutionary
cooling  sequences   of  white   dwarfs  descending   from  metal-rich
progenitors,  hence with  high  $^{22}$Ne abundances.   In summary,  a
realistic assessment of the delay caused by $^{22}$Ne sedimentation in
the  cooling   of  white  dwarfs  resulting   from  Solar  metallicity
progenitors was lacking.  Although the role of $^{22}$Ne sedimentation
becomes  less  relevant   when  the  metal  content   of  white  dwarf
progenitors  is smaller,  the  effect of  $^{22}$Ne  diffusion on  the
cooling of white  dwarfs with solar metallicity might  not be entirely
negligible \citep{2010ApJ...719..612A}.   In this context, the  aim of
this paper was  to provide a grid of cooling  sequences for such white
dwarfs  incorporating  for the  first  time  the effect  of  $^{22}$Ne
sedimentation.  To this end, we  computed the full evolution of 0.528,
0.576, 0.657, and $0.833\, M_{\sun}$ white dwarf models resulting from
the complete evolution of progenitor stars of 1.0, 1.5, 3.0 and $4.0\,
M_{\sun}$ from the ZAMS all the way through the phases of hydrogen and
helium core burning to the thermally  pulsing AGB phase.  In this way,
our  white  dwarf  cooling  sequences  incorporate  realistic  initial
$^{22}$Ne  profiles as  dictated  by nuclear  burning  history of  the
progenitor  star.  In  addition, we  computed $^{22}$Ne  sedimentation
using the  most recent  and reliable $^{22}$Ne  diffusion coefficients
\citep{2010PhRvE..82f6401H}.  Our computations  also take into account
all of  the relevant energy  sources, including latent heat  and phase
separation during crystallization.

We  found   that  $^{22}$Ne  sedimentation  leads   to  a  substantial
lengthening of  the evolutionary  times at  low luminosities  of white
dwarfs resulting from Solar metallicity progenitors. In particular, at
$\log{(L/L\sun)} \approx -4.5$, time delays range from 0.7 to 1.2 Gyr,
depending on the  stellar mass. These delays in the  cooling times are
non  negligible whatsoever.  In fact,  they are  much longer  than the
uncertainties   in  white   dwarf   cooling  times   due  to   current
uncertainties in the stellar microphysics.

We  extend the  scope  of the  paper by  investigating  the impact  of
$^{22}$Ne  sedimentation on  the adiabatic  pulsational properties  of
ZZ~Ceti models.   To this  end, we performed  a pulsation  analysis of
nonradial  $g$ modes.   By the  time  evolution has  proceeded to  the
ZZ~Ceti domain,  element diffusion  has notably altered  the $^{22}$Ne
distribution in the inner regions  of the star.  This has consequences
for  the Brunt-V\"ais\"al\"a  frequency. In  particular, for  low-mass
ZZ~Ceti models we find that the period differences in the range of the
periods observed  in ZZ~Ceti stars  ($100 -  1500$~s) are up  to $\sim
11$~s.   The rate of
period change in low-mass ZZ~Ceti models are not affected by $^{22}$Ne
sedimentation process.  The situation is different in the case of more
massive  stars, for  which $^{22}$Ne  sedimentation becomes  much more
relevant.  Here, we  find that differences in  the theoretical periods
that  result from  including $^{22}$Ne  sedimentation reach  values as
high  as  $\sim  47$~s. The forward period spacings and
the rate of period change are also affected.

We  conclude that  $^{22}$Ne  sedimentation is  a  relevant source  of
energy for white dwarfs  resulting from Solar metallicity progenitors,
that  should be  taken into  account in  the computation  of realistic
cooling sequences for  these stars, as well as in  attempts to perform
precise  asteroseismology of  ZZ~Ceti  stars.  In  particular, in  the
light of our findings,  $^{22}$Ne sedimentation induces non-negligible
changes in  the pulsation periods  and the period spacings  of ZZ~Ceti
stars.  Therefore,  new asteroseismological analysis of  ZZ~Ceti stars
should   be  done   using  stellar   models  that   include  $^{22}$Ne
sedimentation.   In particular,  a re-analysis  on G117$-$B15A  -- the
most well-studied ZZ~Ceti  star -- could help to find  out whether its
main  period (at  $\sim 215$~s),  for which  it has  been possible  to
measure the  secular rate of change,  is indeed a mode  trapped in the
stellar envelope  or not. This is  a crucial aspect in  the context of
the derivation  of the mass of  the axion from pulsating  white dwarfs
\cite{2010A&A...512A..86I,2012MNRAS.424.2792C}.

\acknowledgments 
 We acknowledge the  valuable
comments of our referee which improved  the original version
of this paper. Part  of this work  was supported by  AGENCIA through
the Programa  de Modernizaci\'on  Tecnol\'ogica BID 1728/OC-AR,  by the
PIP   112-200801-00940   grant   from  CONICET,   by   MINECO   grants
AYA2014-59084-P  and  ESP2014-56003-R,  by the  European  Union  FEDER
funds,  and by  grants 2014SGR-1458  and 2014SGR-0038  (Generalitat of
Catalunya).   This research  has made  use of  NASA Astrophysics  Data
System.


\end{document}